\documentclass[conference]{IEEEtran}
\usepackage{amssymb,amsfonts}
\usepackage{algorithmic}
\usepackage{graphicx}
\usepackage{textcomp}
\usepackage{bm}
\usepackage{xcolor}
\usepackage{amsmath}
\usepackage{times}
\usepackage{tikz}
\usepackage{pgfplots}
\usepackage{pgfplotstable}

\pgfplotsset{compat=1.9,
	tick label style={font=\footnotesize},
	label style={font=\normalsize},
	legend style={font=\footnotesize,rounded corners=4pt},
}

\usetikzlibrary{calc, hobby}
\usetikzlibrary{decorations.pathreplacing,decorations.markings,shapes.geometric}
\usetikzlibrary{positioning,shapes,arrows,decorations.markings,calc}
\definecolor{myblue}{RGB}{0,114,189}
\definecolor{myorange}{RGB}{217,83,25}
\definecolor{myyellow}{RGB}{237,177,32 }
\definecolor{mypurple}{RGB}{126,47,142}
\definecolor{mygreen}{RGB}{119,172,48}
\definecolor{mylightblue}{RGB}{77,190,238}
\definecolor{myred}{RGB}{162,20,47}
\definecolor{mylimegreen}{RGB}{170,220,50}

\newcommand{\Fcar}[1]{%
	\begin{tikzpicture}[every node/.append style={rectangle,minimum width=0pt}]
	
\draw[black, thick,fill=cyan!50!white] 
(-1.35,1.4) to[out=-160,in=20] 
(-1.58,1.3) to[out=-170,in=130] 
(-1.65,1.05) to[out=20,in=-160] 
(-1.25,1.15) ;

\draw[black, thick,fill=cyan!50!white] 
(1.35,1.4) to[out=-20,in=160] 
(1.58,1.3) to[out=-10,in=50] 
(1.65,1.05) to[out=160,in=-20] 
(1.25,1.15) ;

	\draw[black, thick,fill=cyan!50!white] 
	(-1.2,-2.5) to[out=115,in=-90] 
	(-1.45,-1.3) to[out=90,in=-95] 
	(-1.35,0) to[out=90,in=-90] 
	(-1.35,2.2) to[out=105,in=-90] 
	(-1.45,3.2) to[out=85,in=-180] 
	(-0.2,4.3) to[out=0,in=180] 
	(0.2,4.3) to[out=0,in=95] 
	(1.45,3.2) to[out=-90,in=75]
	(1.35,2.2) to[out=-90,in=90] 
	(1.35,0) to[out=-90,in=95] 
	(1.45,-1.3) to[out=-90,in=65]
	(1.2,-2.5) to[out=-170,in=-10]
	(-1.2,-2.5) ;

\draw[black, thick,fill=black] 
(-1.15,3.85) to[out=10,in=-135] 
(-0.7,4) to[out=60,in=-95] 
(-0.6,4.23) to[out=-150,in=70] 
(-1.15,3.85);

\draw[black, thick,fill=black!80!white] 
(1.15,3.85) to[out=-170,in=-45] 
(0.7,4) to[out=120,in=-85] 
(0.6,4.23) to[out=-30,in=110] 
(1.15,3.85);

\draw[black!70!white, thin ] 
(-1.05,2.55) to [out=30,in=150] 
(1.05,2.55);

\draw[black!70!white, thin] 
(-1.05,2.9) to [out=20,in=-100] 
(-0.35,4);

\draw[black!70!white, thin] 
(1.05,2.9) to [out=160,in=-80] 
(0.35,4);

	\draw[black, thick,fill=black!80!white] 
	(0,1.2) to[out=180,in=0] 
	(-0.75,1.2) to[out=140,in=-80] 
	(-1.05,2) to[out=80,in=100] 
	(1.05,2) to[out=-100,in=40] 
	(0.75,1.2) to[out=180,in=0] 
	(0,1.2) ;
	
	\draw[black, thick,fill=black!80!white] 
	(0,-1) to[out=180,in=0] 
	(-0.7,-1) to[out=-150,in=100] 
	(-0.65,-1.7) to[out=-60,in=-120] 
	(0.65,-1.7) to[out=80,in=-30] 
	(0.7,-1) to[out=180,in=0] 
	(0,-1) ;
	
	\draw[black, thick,fill=black!80!white] 
	(-1.12,1.4) to[out=-90,in=88] 
	(-1.15,-0.35) to[out=-93,in=120] 
	(-0.95,-0.85) to[out=85,in=-90] 
	(-0.85,0.1) to[out=92,in=-70] 
	(-1.12,1.4) ;
	
	\draw[black, thick,fill=black!80!white] 
	(1.12,1.4) to[out=-90,in=92] 
	(1.15,-0.35) to[out=-87,in=60] 
	(0.95,-0.85) to[out=95,in=-90] 
	(0.85,0.1) to[out=88,in=-110] 
	(1.12,1.4) ;
	
\node[draw,shape=rectangle,fill=white,minimum width=1.2cm,minimum height=0.1cm, outer sep=0pt,inner sep=3pt] at (0,4.1) (ula) {};
	
	\node[fill,circle,inner sep=1.5pt] at (-.9,4.1) (Rx1) {};
	\node[fill,circle,inner sep=1.5pt] at (-.4,4.1) (Rx1) {};
	\node[fill,circle,inner sep=.5pt] at (.05,4.1) (Rx1) {};
	\node[fill,circle,inner sep=.5pt] at (.25,4.1) (Rx1) {};
	\node[fill,circle,inner sep=.5pt] at (.45,4.1) (Rx1) {};
	\node[fill,circle,inner sep=1.5pt] at (.9,4.1) (Rx1) {};

	\end{tikzpicture}
}

\newcommand{\Rcar}[1]{%
	\begin{tikzpicture}[every node/.append style={rectangle,minimum width=0pt}]
	
\draw[black, thick,fill=green!40!white] 
(-1.35,1.4) to[out=-160,in=20] 
(-1.58,1.3) to[out=-170,in=130] 
(-1.65,1.05) to[out=20,in=-160] 
(-1.25,1.15) ;

\draw[black, thick,fill=green!40!white] 
(1.35,1.4) to[out=-20,in=160] 
(1.58,1.3) to[out=-10,in=50] 
(1.65,1.05) to[out=160,in=-20] 
(1.25,1.15) ;

\draw[black, thick,fill=green!40!white] 
(-1.2,-2.5) to[out=115,in=-90] 
(-1.45,-1.3) to[out=90,in=-95] 
(-1.35,0) to[out=90,in=-90] 
(-1.35,2.2) to[out=105,in=-90] 
(-1.45,3.2) to[out=85,in=-180] 
(-0.2,4.3) to[out=0,in=180] 
(0.2,4.3) to[out=0,in=95] 
(1.45,3.2) to[out=-90,in=75]
(1.35,2.2) to[out=-90,in=90] 
(1.35,0) to[out=-90,in=95] 
(1.45,-1.3) to[out=-90,in=65]
(1.2,-2.5) to[out=-170,in=-10]
(-1.2,-2.5) ;

\draw[black, thick,fill=black] 
(-1.15,3.85) to[out=10,in=-135] 
(-0.7,4) to[out=60,in=-95] 
(-0.6,4.23) to[out=-150,in=70] 
(-1.15,3.85);

\draw[black, thick,fill=black!80!white] 
(1.15,3.85) to[out=-170,in=-45] 
(0.7,4) to[out=120,in=-85] 
(0.6,4.23) to[out=-30,in=110] 
(1.15,3.85);

\draw[black!70!white, thin ] 
(-1.05,2.55) to [out=30,in=150] 
(1.05,2.55);

\draw[black!70!white, thin] 
(-1.05,2.9) to [out=20,in=-100] 
(-0.35,4);

\draw[black!70!white, thin] 
(1.05,2.9) to [out=160,in=-80] 
(0.35,4);

\draw[black, thick,fill=black!80!white] 
(0,1.2) to[out=180,in=0] 
(-0.75,1.2) to[out=140,in=-80] 
(-1.05,2) to[out=80,in=100] 
(1.05,2) to[out=-100,in=40] 
(0.75,1.2) to[out=180,in=0] 
(0,1.2) ;

\draw[black, thick,fill=black!80!white] 
(0,-1) to[out=180,in=0] 
(-0.7,-1) to[out=-150,in=100] 
(-0.65,-1.7) to[out=-60,in=-120] 
(0.65,-1.7) to[out=80,in=-30] 
(0.7,-1) to[out=180,in=0] 
(0,-1) ;

\draw[black, thick,fill=black!80!white] 
(-1.12,1.4) to[out=-90,in=88] 
(-1.15,-0.35) to[out=-93,in=120] 
(-0.95,-0.85) to[out=85,in=-90] 
(-0.85,0.1) to[out=92,in=-70] 
(-1.12,1.4) ;

\draw[black, thick,fill=black!80!white] 
(1.12,1.4) to[out=-90,in=92] 
(1.15,-0.35) to[out=-87,in=60] 
(0.95,-0.85) to[out=95,in=-90] 
(0.85,0.1) to[out=88,in=-110] 
(1.12,1.4) ;

\node[draw,shape=rectangle,fill=white,minimum width=1.2cm,minimum height=0.1cm, outer sep=0pt,inner sep=3pt] at (0,-2.5) (ula) {};
	
\node[fill,circle,inner sep=1.5pt] at (-.9,-2.5) (Rx1) {};
	\node[fill,circle,inner sep=1.5pt] at (-.4,-2.5) (Rx1) {};
	\node[fill,circle,inner sep=.5pt] at (.05,-2.5) (Rx1) {};
	\node[fill,circle,inner sep=.5pt] at (.25,-2.5) (Rx1) {};
	\node[fill,circle,inner sep=.5pt] at (.45,-2.5) (Rx1) {};
	\node[fill,circle,inner sep=1.5pt] at (.9,-2.5) (Rx1) {};

	\end{tikzpicture}
}

\newcommand{\he}{\operatorname{H}}
\newcommand{\tr}{\operatorname{T}}
\newcommand{\jj}{\operatorname{j}}

\newcommand{\m}[1]{\mathbf{#1}}

\usepackage{textcomp}
\newcommand\copyrighttext{%
	\footnotesize \textcopyright 2018 IEEE. Personal use of this material is permitted. Permission from IEEE must be obtained for all other uses, in any current or future media, including reprinting/republishing this material for advertising or promotional purposes, creating new collective works, for resale or redistribution to servers or lists, or reuse of any copyrighted component of this work in other works.}

\newcommand\copyrightnotice{%
	\begin{tikzpicture}[remember picture,overlay]
	\node[anchor=south,yshift=10pt] at (current page.south) {{\parbox{\dimexpr\textwidth-\fboxsep-\fboxrule\relax}{\copyrighttext}}};
	\end{tikzpicture}%
}

\newcommand\conferenceinfotext{%
	\footnotesize This paper has been published at the 2018 IEEE Vehicular Technology Conference - VTC Fall 2018, where it was selected as the \emph{IEEE VTC 2018-Fall Conference's Best Paper}.}

\newcommand\conferenceinfonotice{%
	\begin{tikzpicture}[remember picture,overlay]
	\node[anchor=south,yshift=-27pt] at (current page.north) {{\parbox{\dimexpr\textwidth-\fboxsep-\fboxrule\relax}{\conferenceinfotext}}};
	\end{tikzpicture}%
}

\begin{document}
%
\title{LOS MIMO Design based on Multiple Optimum Antenna Separations}

\author{\IEEEauthorblockN{Mario H. Casta\~{n}eda Garcia$^{*}$, Marcin Iwanow$^{*,\dagger}$ and Richard A. Stirling-Gallacher$^{*}$}
	\IEEEauthorblockA{
		$^{*}$\textit{Munich Research Center}, \textit{Huawei Technologies D\"usseldorf GmbH,  Munich, Germany} \\
		$^{\dagger}$\textit{Associate Institute for Signal Processing, Technische Universit\"at M\"unchen,  Munich, Germany}  \\		
		\texttt{\{mario.castaneda,iwanow.marcin,richard.sg\}}@huawei.com}
}



\maketitle
\IEEEoverridecommandlockouts
\copyrightnotice
\conferenceinfonotice

\vspace*{-0.35cm}
\begin{abstract}	
The use of multiple antennas in a transmit and receive antenna array for MIMO wireless communication allows the spatial degrees of freedom in rich scattering environments to be exploited. However, for \emph{line-of-sight} (LOS) MIMO channels  with \emph{uniform linear arrays} (ULAs) at the transmitter and receiver, the antenna separations at the transmit and receive array need to be optimized to maximize the spatial degrees of freedom and the channel capacity. In this paper, we first revisit the derivation of the optimum antenna separation at the transmit and receive ULAs in a LOS MIMO system, and provide the \emph{general} expression for the optimum antenna separation product, which consists of \emph{multiple} solutions. Although only the solution corresponding to the smallest antenna separation product is usually considered in the literature, we exploit the multiple solutions for a LOS MIMO design over a \emph{range} of distances between the transmitter and  receiver. In particular, we consider the LOS MIMO design  in a \emph{vehicle-to-vehicle} (V2V) communication scenario, over a range of distances between  the transmit and receive vehicle.
\end{abstract}


%
\IEEEpeerreviewmaketitle

\section{Introduction}
The spatial degrees of freedom offered by a MIMO system with a transmit and receive antenna array can be exploited in the presence of a rich scattering environment. However, in LOS MIMO channels with little or no scattering, the channel responses can become highly correlated, leading to a MIMO channel of rank 1. Nevertheless, with a proper placement of the  antennas in the arrays \cite{Driessen99,Gesbert02,Haustein03}, the channel capacity and rank of the LOS MIMO channel can be maximized.

With ULAs at the transmitter and receiver, the best antenna placement is obtained by optimizing the separation between the antennas in the transmit and receive arrays. Although for ULAs there are multiple solutions \cite{Bohagen05,Sarris07} for the optimum antenna separation product, i.e. the product of the antenna separation at the transmit and receive array, the one corresponding to the \emph{smallest} antenna separation product is usually considered, as this leads to the smallest arrays \cite{Bohagen05}.

The previous cited works consider a fixed distance  between the transmit and receive array. However, for many applications, a LOS MIMO channel needs to be designed for a \emph{range} of distances between the transmitter and receiver. Since the optimum antenna separation depends on the distance between the transmitter and receiver, there is a performance degradation when the distance is varied for a given antenna placement of the transmit and receive arrays. To reduce the sensitivity to distance variations between the transmitter and receiver, non-uniform linear arrays  have been proposed \cite{Torkildson09,Zhou13}. The optimum antenna placement in such cases was found using an exhaustive search, with the
 aim of maximizing the range where a minimum  condition number or capacity can be guaranteed.

In this paper, we first revisit the derivation of the optimum antenna separation for LOS MIMO systems with ULAs at the transmitter and receiver. In contrast to prior work, we provide the \emph{general} expression for the optimum antenna separation product, which consists of multiple solutions. In addition, we propose to use the multiple solutions for the LOS MIMO design over a range of distances. In particular, we consider the LOS MIMO design for a V2V communication scenario over a range of distances between the transmit and receive vehicle. Although the optimum antenna placement can not be met at all distances, we exploit the fact that some antenna separations are optimum at \emph{several} distances. We show that \emph{larger} antenna separations can  be beneficial in certain cases.

This paper is organized as follows. Section~\ref{Sec:ChMod} introduces the LOS MIMO channel model. The optimum antenna separation is derived in Section~\ref{Sec:OptAntSep}. The V2V scenario is described in Section~\ref{Sec:V2V}, where numerical results for the LOS MIMO design are presented. We conclude the paper with Section~\ref{Sec:Con}. 

\section{LOS MIMO Channel Model}
\label{Sec:ChMod}

In this paper, we use lower case and capital boldface letters to denote vectors and matrices, respectively. In addition,  $(\bullet)^{\tr}$ and $(\bullet)^{\he}$ denote the transpose and conjugate transpose, respectively. The cardinality of the set $\mathcal{P}$ is denoted by $|\mathcal{P}|$. 

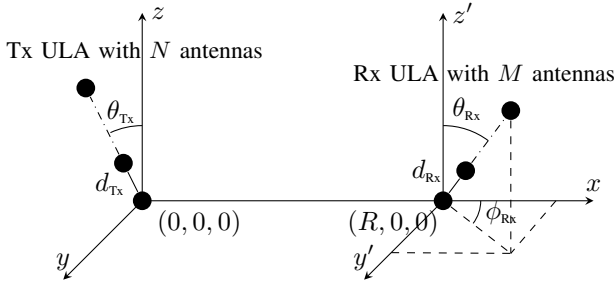
\begin{figure}[!ht]
	\begin{center}
		\begin{tikzpicture}[x=0.5cm,y=0.5cm,z=0.3cm,>=stealth]
		\draw[->] (xyz cs:x=0) -- (xyz cs:x=12) node[above] {$x$};
		\draw[->] (xyz cs:y=0) -- (xyz cs:y=5) node[right] {$z$};
		\draw[->] (xyz cs:z=0) -- (xyz cs:z=-3.5) node[above] {$y$};
		\node[fill,circle,inner sep=2.5pt] at (0,0) (Tx1) {};
		\node[fill,circle,inner sep=2.5pt] at (-0.5,1) (Tx2) {};
		\node[fill,circle,inner sep=2.5pt] at (-1.5,3) (TxN) {};
		\draw[-] (Tx1) node[below] {\hspace{8ex} {  $(0,0,0)$}} -- (Tx2) 
		 node[below] {\hspace{-3ex} $d_{\text{\tiny Tx}}$};
		\draw[dashdotted] (Tx2) -- (TxN)  {};
		\draw (0,2) arc (90:115:2) node[above] {\hspace{1ex} $\theta_{\text{\tiny Tx}}$};
		\node at (-0.2,4) () {\small  Tx ULA with $N$ antennas};
		\draw[->] (xyz cs:x=8,y=0) -- (xyz cs:x=8,y=5) node[right] {$z^{\prime}$};
		\draw[->] (xyz cs:x=8,z=0) -- (xyz cs:x=8,z=-3.5) node[above] {$y^{\prime}$};
		\node[fill,circle,inner sep=2.5pt] at (8,0) (Rx1) {};
		\node[fill,circle,inner sep=2.5pt] at (8.6,0.8) (Rx2) {};
		\node[fill,circle,inner sep=2.5pt] at (9.8,2.4) (RxN) {};
		\draw[-] (Rx1) -- (Rx2)  node[left] { \hspace{-7ex} $d_{\text{\tiny Rx}}$};
		\draw[dashdotted] (Rx2) -- (RxN)   {};
		\node at (9.1,3.4) () {\small   Rx ULA with $M$ antennas };
		\draw[dashed] (RxN) -- (xyz cs:x=9.8,y=-1.4) ;
		\draw[dashed] (Rx1)  node[below] {\hspace{-10ex} {  $(R,0,0)$}} -- (xyz cs:x=9.8,y=-1.4) ;
		\draw[dashed] (xyz cs:x=8,z=-2.3) -- (xyz cs:x=9.8,y=-1.4) ;
		\draw[dashed] (xyz cs:x=11) -- (xyz cs:x=9.8,y=-1.4) ;
		\draw (8,2) arc (90:53:2)  node[midway, above] {$\theta_{\text{\tiny Rx}}$};
		\draw (9,0) arc (0:-37:1) node[midway,right] {\hspace{-1ex} $\phi_{\text{\tiny Rx}}$};
		%
		
		\end{tikzpicture}
	\end{center}
	\vspace{-1.5ex} 
	\caption{LOS MIMO channel with a transmit and receive ULA}
	\label{fig:LOSMIMO}
\end{figure}

We consider a MIMO  channel  with a pure LOS  between a transmitter and a receiver consisting of a ULA with $N > 1$ and $M > 1$ antennas, respectively. The antenna separation at the \emph{transmit} (Tx) and \emph{receive} (Rx) ULA is $d_{\text{\tiny Tx}}$  and $d_{\text{\tiny Rx}}$, respectively. The distance between the first antenna of the Tx ULA, placed at the origin, and the first antenna of the Rx ULA is given by $R$ as shown in Fig.~\ref{fig:LOSMIMO}. With the Tx array placed on the $xz$-plane, we assume an arbitrary orientation of the arrays given by the angles $\theta_{\text{\tiny Tx}}$, $\theta_{\text{\tiny Rx}}$ and $\phi_{\text{\tiny Rx}}$ as shown in Fig.~\ref{fig:LOSMIMO}, where $0 \le \theta_{\text{\tiny Tx}} \le \frac{\pi}{2}$ and $0 \le \theta_{\text{\tiny Rx}} \le \frac{\pi}{2}$. The carrier frequency and  wavelength of the signal are given by $f_{\text{c}}$ and $\lambda$, respectively.

The \emph{normalized} channel matrix for the LOS MIMO system is denoted as 
\begin{align}
\m{H} = \left[ \begin{array}{cccc}
\m{h}_{1} &  \m{h}_{2} & \cdots & \m{h}_{N}
\end{array} \right] \in \mathbb{C}^{M \times N},
\label{def_H}
\end{align}
where the $n$-th column of $\m{H}$, i.e. $\m{h}_{n}$, corresponds to the channel vector from the $n$-th antenna at the Tx array to the $M$ antennas at the Rx antenna array. With the path loss included in the receive SNR, the \emph{normalized} channel vector $\m{h}_{n} \in \mathbb{C}^{M}$ is determined with ray tracing, i.e. with the spherical wave model instead of the planar wave assumption, and is given as:
\begin{align}
\m{h}_{n} = \left[ \begin{array}{ccc}
\text{exp}\left( \jj 2\pi \frac{r_{1,n}}{\lambda}\right), & \cdots, &  \text{exp}\left( \jj 2\pi \frac{r_{M,n}}{\lambda}\right)
\end{array} \right]^{\tr}.
\label{def_hn}
\end{align}
where 
$r_{m,n}$ corresponds to the path length between the $n$-th Tx antenna  and the $m$-th Rx antenna, for $n=1,\ldots,N$ and $m=1,\ldots,M$, respectively. The path length $r_{m,n}$ can be obtained from the coordinates $(x^{\text{\tiny Tx}}_n,y^{\text{\tiny Tx}}_n,z^{\text{\tiny Tx}}_n)$ of the $n$-th Tx antenna and the coordinates $(x^{\text{\tiny Rx}}_m,y^{\text{\tiny Rx}}_m,z^{\text{\tiny Rx}}_m)$ of the $m$-th Rx antenna, which from Fig.~\ref{fig:LOSMIMO} are given by 
\begin{align*}
\begin{array}{ll}
\text{$n$-th Tx ant.}\!: & x^{\text{\tiny Tx}}_n =-(n-1) d_{\text{\tiny Tx}} \sin \theta_{\text{\tiny Tx}},   \\
& y^{\text{\tiny Tx}}_n = 0, \quad \quad z^{\text{\tiny Rx}}_m = (n-1) d_{\text{\tiny Tx}} \cos \theta_{\text{\tiny Tx}}  \\
\text{$m$-th Rx ant.}\!: & x^{\text{\tiny Rx}}_m = R \!+\!(m\!-\!1) d_{\text{\tiny Rx}} \sin \theta_{\text{\tiny Rx}} \cos \phi_{\text{\tiny Rx}}, \\
&\hspace{-12ex} y^{\text{\tiny Rx}}_m \!=\! (m\!-\!1) d_{\text{\tiny Rx}} \sin \theta_{\text{\tiny Rx}} \sin \phi_{\text{\tiny Rx}}, 
 \quad\,  z^{\text{\tiny Rx}}_m \!=\!(m\!-\!1) d_{\text{\tiny Rx}} \cos \theta_{\text{\tiny Rx}}.
\end{array}
\end{align*}
With the above coordinates,  $r_{m,n}$ can be determined as follows
\begin{align}
r_{m,n} &\boldsymbol{=} \left(\left(x^{\text{\tiny Rx}}_m-x^{\text{\tiny Tx}}_n\right)^{2} + \left(y^{\text{\tiny Rx}}_m-y^{\text{\tiny Tx}}_n\right)^{2} + \left(z^{\text{\tiny Rx}}_m-z^{\text{\tiny Tx}}_n\right)^{2} \right)^{\frac{1}{2}} \nonumber \\
&\hspace{-5ex} \boldsymbol{=}\! \! \left(\!\left(R \!+\!(m\!-\!1) d_{\text{\tiny Rx}} \sin \theta_{\text{\tiny Rx}} \! \cos \phi_{\text{\tiny Rx}}\!\! +\! (n\!-\!1) d_{\text{\tiny Tx}} \sin \theta_{\text{\tiny Tx}} \!\right)^{2} \!\!\! +\! \big(\!(m\!-\!1) \times \right. \nonumber \\
& \hspace{-5ex} \left. d_{\text{\tiny Rx}}  \sin \theta_{\text{\tiny Rx}} \sin \phi_{\text{\tiny Rx}}\big)^{2} \!\! + \!
 \left((m\!-\!1) d_{\text{\tiny Rx}} \! \cos  \theta_{\text{\tiny Rx}} \! - \!(n\!-\!1) d_{\text{\tiny Tx}}  \cos \theta_{\text{\tiny Tx}} \right)^{2}  \right)^{\!\frac{1}{2}} \nonumber \\
&\hspace{-5ex} \boldsymbol{=} \Big(R \!+\!(m\!-\!1) d_{\text{\tiny Rx}} \sin \theta_{\text{\tiny Rx}} \cos \phi_{\text{\tiny Rx}} + (n-1) d_{\text{\tiny Tx}} \sin \theta_{\text{\tiny Tx}} \Big) \Big( 1 +  \nonumber \\
& \hspace{-6ex}  \frac{\left( \! (m\!-\!1) d_{\text{\tiny Rx}} \sin \! \theta_{\text{\tiny Rx}} \sin \!\phi_{\text{\tiny Rx}}\right)^{2} \!\!\! + \!
 \left(\!(m\!-\!1) d_{\text{\tiny Rx}} \cos\! \theta_{\text{\tiny Rx}}\! \! - \!(n\!-\!1) d_{\text{\tiny Tx}} \!\cos \! \theta_{\text{\tiny Tx}}\!\right)^{\!2}}{\left(R \!+\!(m\!-\!1) d_{\text{\tiny Rx}} \sin \theta_{\text{\tiny Rx}} \cos \phi_{\text{\tiny Rx}} + (n-1) d_{\text{\tiny Tx}} \sin \theta_{\text{\tiny Tx}}\right)^{2}} \!\bigg)^{\!\!\!\frac{1}{2}} \nonumber \\ 
& \hspace{-5ex}\boldsymbol{\approx} R \!+\!(m\!-\!1) d_{\text{\tiny Rx}} \sin \theta_{\text{\tiny Rx}} \cos \phi_{\text{\tiny Rx}} + (n-1) d_{\text{\tiny Tx}} \sin \theta_{\text{\tiny Tx}} + \nonumber \\
& \hspace{-6ex} \frac{\left(\! (m\!-\!1) d_{\text{\tiny Rx}}\! \sin \theta_{\text{\tiny Rx}} \! \sin \phi_{\text{\tiny Rx}}\right)^{2} \!\!\! + \!
		\left(\!(m\!-\!1) d_{\text{\tiny Rx}} \!\cos  \theta_{\text{\tiny Rx}} \! - \!(n\!-\!1) d_{\text{\tiny Tx}}\! \cos  \theta_{\text{\tiny Tx}}\right)^{2}}{2 R},	\label{def_rmn}
\end{align}
where the last step results from the first order approximation of the Taylor series of $\sqrt{1+a}$ with $a \ll 1$, i.e. $\sqrt{1+a}\approx 1 + \frac{a}{2}$, and from  $R \approx R+ (m\!-\!1) d_{\text{\tiny Rx}} \sin \theta_{\text{\tiny Rx}} \cos \phi_{\text{\tiny Rx}} + (n-1) d_{\text{\tiny Tx}} \sin \theta_{\text{\tiny Tx}}$ in the denominator of the argument of the square root, where both approximations hold if the distance $R$ between the transmitter and receiver is much larger than Tx and Rx array dimensions.

\section{Optimum Antenna Separation}
\label{Sec:OptAntSep}

Consider the case when $N \le M$, such that\footnote{The case $N > M$ can be derived in a similar  manner by simply interchanging the tranmsitter and the receiver.} $\text{rank}\left(\m{H}\right) \le N$. As discussed in \cite{Bohagen05}, the capacity of the LOS MIMO system at high SNR is maximized if $\m{H}^{\text{H}}\m{H} = M \m{1}_{N}$, i.e. if the columns of $\m{H}$ are orthogonal. For this case, $\m{H}$ achieves the maximum rank of $N$ and the $N$ eigenvalues of $\m{H}^{\text{H}}\m{H}$ are all equal to $M$, as $\text{tr}\left(\m{H}^{\text{H}}\m{H}\right)=MN$.

\subsection{Solution of the Orthogonality Condition}

In order to design the channel matrix $\m{H}$ of the LOS MIMO system to have orthogonal columns, from (\ref{def_H}) we need to have
\begin{align}
\m{h}_{k}^{\he} \m{h}_{l} = 0, \quad \quad \text{for} \quad k \ne l; \quad  k,l=1,\cdots,N.
\label{orth_cond}
\end{align} 
Using (\ref{def_hn}), we can write 
\begin{align}
\m{h}_{k}^{\he} \m{h}_{l} 
&= \sum_{m=1}^{M} \text{exp}\left(\jj 2\pi \frac{r_{m,l}-r_{m,k}}{\lambda} \right) \nonumber \\
&\!\!\!\!\overset{\text{(a)}}{\approx} \! \sum_{m=1}^{M} \! \text{exp}\left(\! \jj  2\pi \! \left(\!\frac{\gamma}{\lambda}\! -\!\frac{d_{\text{\tiny Tx}} d_{\text{\tiny Rx}} \cos{\theta_{\text{\tiny Tx}}} \cos{\theta_{\text{\tiny Rx}}}}{\lambda R} (l\!-\!k) (m\!-\!1) \!\right) \!\right) \nonumber \\
&\!\!\!\!\overset{\text{(b)}}{=} \Gamma  \cdot \sum_{m^{\prime}=0}^{M-1} \!\! \text{exp}\left(\jj 2\pi \frac{d_{\text{\tiny Tx}} d_{\text{\tiny Rx}} \cos{\theta_{\text{\tiny Tx}}} \cos{\theta_{\text{\tiny Rx}}}}{\lambda R} (k-l) m^{\prime} \right) \nonumber \\
&\!\!\!\!\overset{\text{(c)}}{=} \Gamma \cdot
\frac{1-\text{exp}\left(\jj 2\pi \frac{d_{\text{\tiny Tx}} d_{\text{\tiny Rx}} \cos{\theta_{\text{\tiny Tx}}} \cos{\theta_{\text{\tiny Rx}}}}{\lambda R} M (k-l)  \right)}{1-\text{exp}\left(\jj 2\pi \frac{d_{\text{\tiny Tx}} d_{\text{\tiny Rx}} \cos{\theta_{\text{\tiny Tx}}} \cos{\theta_{\text{\tiny Rx}}}}{\lambda R} (k-l) \right)},
\label{inner_prod}
\end{align}
where step (a) results from 
\begin{align}
r_{m,l}-r_{m,k} \approx & 
\, \gamma -  \frac{d_{\text{\tiny Tx}} d_{\text{\tiny Rx}}  \cos \theta_{\text{\tiny Tx}} \cos  \theta_{\text{\tiny Rx}}}{R} (l\!-\!k) (m\!-\!1),
\end{align}
which follows from using the approximation (\ref{def_rmn}) for $r_{m,n}$, and where 
$\gamma \!=\! (l\!-\!k) d_{\text{\tiny Tx}} \sin \theta_{\text{\tiny Tx}} - \frac{ \left((l-1)^{2}\! - (k-1)^{2} \right)d^{2}_{\text{\tiny Tx}} \cos^{2} \theta_{\text{\tiny Tx}}}{2 R}$. For step (b), we use the substitutions  $m^{\prime}=m\!-\!1$ and $\Gamma\! \!= \!\text{exp}\left(\jj  2\pi \! \frac{\gamma}{\lambda}\right)$, with $\Gamma$ being  independent of $m^{\prime}$.  For step (c), we employ the expression for the finite sum of a geometric series for $w\ne 1$:
\begin{align}
\sum_{m^{\prime}=0}^{M-1}  w^{m^{\prime}} =   \frac{1 - w^M}{1-w} ,
\end{align}
with $w = \text{exp}\left(\jj 2\pi \frac{d_{\text{\tiny Tx}} d_{\text{\tiny Rx}} \cos{\theta_{\text{\tiny Tx}}} \cos{\theta_{\text{\tiny Rx}}}}{\lambda R} (k-l) \right) $.

Given that $\m{h}_{k}^{\he} \m{h}_{l}$ depends on $(k-l)$, as observed from (\ref{inner_prod}), and  that $|\m{h}_{k}^{\he} \m{h}_{l}|=|\m{h}_{l}^{\he} \m{h}_{k}|$, the conditions given in (\ref{orth_cond}) required to have orthogonal columns of $\m{H}$ are equivalent to 
\begin{align}
\m{h}_{k}^{\he} \m{h}_{l} = 0, \quad \quad \text{for} \quad (k -l)=1,\cdots,N-1.
\label{orth_cond_0}
\end{align} 
From (\ref{inner_prod}) and as $\Gamma \ne 0$, the  {equivalent} orthogonality  conditions in  \eqref{orth_cond_0} are fulfilled\footnote{Due to the approximation (\ref{def_rmn}) for $r_{m,n}$, (\ref{orth_cond_0}) can only be fulfilled approximately with (\ref{ortho_cond_1}). As the error introduced with (\ref{def_rmn}) is negligible for practical systems \cite{Sarris07}, we assume in the following that (\ref{ortho_cond_1}) can be met with equality.} if
\begin{align}
\frac{1-\text{e}^{\jj 2\pi \delta M  q}}{1-\text{e}^{\jj 2\pi \delta q }}  =0,  \quad    \forall \, q\in\{1, 2,\! \cdots\!, N\!-\!1\}, 
 \label{ortho_cond_1}
\end{align}
where we introduce $q=k-l$ and define
\begin{align}
\delta \overset{\Delta}{=}
 \frac{d_{\text{\tiny Tx}} d_{\text{\tiny Rx}} \cos{\theta_{\text{\tiny Tx}}} \cos{\theta_{\text{\tiny Rx}}}}{\lambda R}.
\label{delta}
\end{align}

Solving \eqref{ortho_cond_1} with respect to $\delta$, allows us to determine the optimum antenna separations $d_{\text{\tiny Tx}}$ and $d_{\text{\tiny Rx}}$ of the Tx and Rx ULAs,  which lead to a channel matrix $\m{H}$ that maximizes the capacity of the LOS MIMO system. 

To satisfy (\ref{ortho_cond_1}), the numerator of the expression in  (\ref{ortho_cond_1}) needs to be zero while the denominator is non-zero, i.e. 
\begin{align}
\text{e}^{\jj 2\pi \delta M  q}=1, \quad    \forall \, q \in \{1, 2,\! \cdots\!, N\!-\!1\},
\label{num_zero_0}
\end{align}
while
\begin{align}
\text{e}^{\jj 2\pi \delta   q} \ne 1, \quad    \forall \, q \in\{1, 2,\! \cdots\!, N\!-\!1\}.
\label{den_zero_0}
\end{align}

As the solution of (\ref{num_zero_0}) for $q=1$ is also a solution of (\ref{num_zero_0}) for $q=2,\cdots,N-1$, the solution of (\ref{num_zero_0}) for all $q$ results from $\text{e}^{\jj 2\pi \delta M }=1$, i.e. the solution of (\ref{num_zero_0}) is
\begin{align}
\delta = \frac{p}{M},  \quad \forall \,\, p \in \mathbb{Z}_{+}, \label{num_zero} 
\end{align}
where $\mathbb{Z}_{+}$ represents the set of positive integers.  The set of negative integers  is excluded from the solution since all the terms in $\delta$ are positive, as can be seen in \eqref{delta}. 

On the other hand, to avoid the denominator of the expression in \eqref{ortho_cond_1} being zero for any value of $q$, from \eqref{den_zero_0} we get 
\begin{align}
&\delta \ne \frac{p_1}{q},  \quad \, \, \forall \, p_1 \in \mathbb{Z}_{+},  \,\,   q\in\{1, 2,\! \cdots\!, N\!-\!1\}. 
\label{den_zero} 
\end{align}

Thus, given \eqref{num_zero} and \eqref{den_zero}, we have that (\ref{ortho_cond_1}) is fulfilled if $\delta = \frac{p}{M}$ for $p \in \mathbb{Z}_{+}$ but excluding the  integers $p$ for which  $\frac{p}{M} = \frac{p_1}{q}$ for $q= 1, 2,\cdots, N\!-\!1$, i.e. when
\begin{align}
\delta = \frac{p}{M},   \quad \forall \, p \in \mathbb{Z}_{+} \setminus \left\{p^{\prime}: p^{\prime} \!=\! \frac{p_1 M}{q}, \!\!\! \begin{array}{c} p_1 \in \mathbb{Z}_{+} , p^{\prime} \in \mathbb{Z}_{+},  \\
q\in\{1, 2,\! \cdots\!, N\!-\!1\}
\end{array}
\!\!\!\! \right\} \!\! .  \label{ortho_sol_0}
\end{align}
Writing $q$ as the product of \emph{any} two (positive integer) factors, i.e. $q=q_1 q_2$,  $p^{\prime}=\frac{p_1 M}{q}$ is an integer if  $\frac{p_1}{q_1}$ and $\frac{M}{q_2}$ are both integers. As there is always a $p_1\in \mathbb{Z}_{+}$ such that $\frac{p_1}{q_1} \in \mathbb{Z}_{+}$, we only need to consider when $\frac{M}{q_2}$ is an integer for any factor $q_2$ of $q$. Given that $q_2 \le q \le N-1$, $\forall q$, we have that $p^{\prime}=\frac{p_1 M}{q}$ is an integer if  $\frac{M}{q}$ is an integer for $q=1, \ldots, N-1$. 
The possible values, in ascending order, of $\frac{M}{q}$ for  $q=1, \ldots, N-1$, are $\frac{M}{N-1}$, $\frac{M}{N-2}$, \ldots $\frac{M}{2}$, $M$, out of which those that are integers (recall that $N \le M$), correspond to the \emph{divisors} of $M$ which are larger than or equal to $\frac{M}{N-1}$. Let us denote the set of divisors of $M$ which satisfy this condition as $\mathcal{D}_M(N)$, i.e.
\begin{align}
\mathcal{D}_{M}(N)= \left\{\nu: \nu \,\, | \, \, M, \,\,\nu \ge  \frac{M}{N-1} \right\} , 
\label{div_set}
\end{align}
where $a \,\, | \, \, b$ means that $a$ is a divisor of $b$. Given \eqref{div_set}, we can  rewrite the solution \eqref{ortho_sol_0} for the orthogonality conditions  as 
\begin{align}
\delta = \frac{p}{M},  \quad  \forall \, p \in \mathbb{Z}_{+} \setminus \, \left\{ 
p^{\prime} \, \nu, \, \, p^{\prime} \in \mathbb{Z}_{+}, \, \nu \in \mathcal{D}_M(N) \right\},
\label{ortho_sol}
\end{align}
i.e. \eqref{ortho_cond_1} is fulfilled if $\delta = \frac{p}{M}$ for the set of positive integers $p$ \emph{excluding} the multiples of divisors of $M$ which are larger than or equal to $\frac{M}{N-1}$.

Prior solutions of (\ref{ortho_cond_1}) provided in the literature, e.g. as in \cite{Sarris07}, include only a subset of the possible integers $p$ given in (\ref{ortho_sol}). In addition, in contrast to prior work, our derived expression (\ref{ortho_sol}) shows  the dependency on $N$, which corresponds to the number of Tx antennas and the $\text{rank}(\m{H})$. We discuss this dependency with two examples: $N\!=\!2$ and $N\!=\!M$. For $N\!=\!2$, $\frac{M}{N-1}=M$ such that from \eqref{div_set}, $\mathcal{D}_M(2)=\{M\}$. On the other hand, for $N\!=\!M$,  $\frac{M}{N-1}=1+\frac{1}{M-1}$ such that $\mathcal{D}_M(M) = \left\{\nu: \nu \,\, | \, \, M, \,\,\nu > 1 \right\}$, i.e. $\mathcal{D}_M(M)$ consists of all the divisors\footnote{If $M$ is prime, $\mathcal{D}_M(M)=\{M\}$ and hence, (\ref{ortho_sol}) is independent of  $N$.} of $M$ \emph{except} $1$. As  $|\mathcal{D}_M(M)| \ge |\mathcal{D}_M(2)|$, we see that in general a larger set of positive integers $p$ are excluded in (\ref{ortho_sol}) when $N=M>2$ compared to when $N=2$. This is a consequence of the fact that the orthogonality conditions in (\ref{ortho_cond_1}) becomes more stringent with increasing $N$: for $N=2$, only two channel vectors need to be orthogonal, whereas for $N=M$, $M$ orthogonal channel vectors need to be designed.

\subsection{Design of LOS MIMO Systems}

Using \eqref{delta}, we rewrite (\ref{ortho_sol})  in terms of the  \emph{antenna separation product} (ASP) \cite{Bohagen05}, i.e. in terms of the product of the antenna separation at the transmitter and receiver
\begin{align}
&\quad  d_{\text{\tiny Tx}} d_{\text{\tiny Rx}} = p \cdot \frac{\lambda R}{M \cos{\theta_{\text{\tiny Tx}}} \cos{\theta_{\text{\tiny Rx}}}},  \label{ortho_sol_asp} 
\\   &\forall \, p \in \mathbb{Z}_{+} \setminus \, \left\{ 
p^{\prime} \, \nu, \, \, p^{\prime} \in \mathbb{Z}_{+}, \, \nu \in \mathcal{D}_M(N) \right\},
\nonumber
\end{align}
for $N \le M$. For $N > M$, the optimum solution for the ASP results from exchanging $N$ with $M$ in the expression above.  

By setting the antenna separations $d_{\text{\tiny Tx}}$ and $d_{\text{\tiny Rx}}$ of the Tx and Rx ULAs according to (\ref{ortho_sol_asp}), the channel matrix $\m{H}$ of the LOS MIMO system can be designed to have orthogonal columns, for a given distance $R$ between the arrays and a given orientation of the arrays. Although multiple solutions for the ASP exist\footnote{Despite infinite solutions, not all solutions fulfill (\ref{orth_cond_0}) in practice. As $p\! \rightarrow \!\infty$, the length of the arrays increase such that the assumption that the distance between the arrays is much larger than the array dimensions becomes invalid.}, only the first solution of \eqref{ortho_sol_asp}, i.e. $p=1$, is usually considered in the literature as this leads to the smallest antenna separations and hence, to the smallest arrays \cite{Bohagen05,Sarris07}. 

However, for certain applications, \emph{other} solutions for the ASP, i.e. $p>1$,  might  be of interest. Take for instance the LOS MIMO design over a \emph{range} of distances between the transmitter and receiver, which is relevant for many applications. As observed in (\ref{ortho_sol_asp}), the optimum antenna separations at the Tx and Rx arrays depends on the \emph{fixed} distance $R$ between the arrays. Thus, varying the distance between the transmitter and receiver with a given optimum antenna separation, leads to a capacity reduction, i.e. reduced $\text{rank}(\m{H})$  or non-equal eigenvalues of $\m{H}^{\text{H}}\m{H}$. To reduce the sensitivity to distance variations, non-uniform linear arrays  have been proposed \cite{Torkildson09,Zhou13}, where the optimum antenna placement is found via an exhaustive search, in order to maximize the range for which a certain metric can be guaranteed. In this paper, we propose the use of ULAs for the LOS MIMO design over a set of distances between the transmitter and receiver, by exploiting the multiple solutions for the ASP given in (\ref{ortho_sol_asp}). In particular, we consider the LOS MIMO design for a V2V link as discussed next.

\section{LOS MIMO Design for V2V} 	
\label{Sec:V2V}

\begin{figure}[!ht]
	\vspace{-1ex} 
	\begin{center}
		\begin{tikzpicture}[x=0.5cm,y=0.5cm,>=stealth,scale=0.65, every node/.style={transform shape}]
		\draw[-,dash pattern=on 8pt off 8pt,line width=1.5pt] (0,-2.2) -- (0,20.2) {};
		\draw[-,line width=1.75pt] (5,-2.5) -- (5,20.5) {};
		\draw[-,line width=1.75pt] (-5,-2.5) -- (-5,20.5) {};
		\draw[<->,loosely dashed] (0,-2.7) -- (5,-2.7) node[align=center, midway,below=5pt] {{  Lane Width} \\ {  3.5 m}};
		\draw[-,dashed] (-1.1,4.4) -- (7.5,4.4) {};
		\draw[-,dashed] (-1.1,13.6) -- (6,13.6) {};
		\draw[<->,loosely dashed] (-0.6,4.4) -- (-0.6,13.6) node[align=center,midway,left] {\large{Longitudinal \, Distance $D$} \\ {\large  between \, cars \hspace{2ex}}};
		\node[] (RxCar) at (2.9,1.1) {\Fcar{}}; 
		\node[] (TxCar) at (2,16.9) {\Rcar{}}; 
		\draw[-,dashed] (2.225,2) -- (0.875,16) {};
		\draw[<->] (2,4.4) -- (1.1,13.6) node[midway,right] {\large  $R$};
		\draw[-,dashed] (0,13.5) -- (6,14.0784) {};
		\draw[-,dashed] (1,4.3) -- (7.5,4.8784) {};
		\draw (7,4.4) arc (0:5:5) node[midway,xshift=9pt] {\hspace{3ex} \large $\theta_{\text{\tiny Rx}}$};
		\draw (5.5,13.6) arc (0:5:5) node[midway,xshift=9pt] {\hspace{3ex} \large $\theta_{\text{\tiny Tx}}$};
		\draw[->] (-6.5,1.5) node[left,align=center] { {\large  Rx Car with} \\ {\large  Rx ULA in}  \\ {\large  front bumper} } -- (2.9,0.5);
		\draw[->] (-6.5,17) node[left,align=center] { {\large  Tx Car with} \\ {\large  Tx ULA in}  \\ {\large  rear bumper} } -- (2,16)  ;
		\draw[decoration={brace,raise=5pt},decorate]	(1.85,4.5) --  node[above=10pt] {\large \hspace{2ex}  $\le L_{\text{\tiny Rx}}$} (3.9,4.5);		
		\draw[decoration={brace,mirror,raise=5pt},decorate]	(.95,13.5) --  node[below=10pt] {\large \hspace{2ex} $\le L_{\text{\tiny Tx}}$} (3.05,13.5);
		\draw[->] (6,11) node[right,align=center] { {\large  Tx ULA with} \\ {\large  $N$ antennas}  } -- (3.05,13.5);
		\draw[->] (6,7) node[right,align=center] { {\large  Rx ULA with} \\  {\large  $M$ antennas} } -- (3.9,4.5);
		

		\end{tikzpicture}
	\end{center}
	\vspace{-3ex} 
	\caption{V2V Scenario with two vehicles within a lane}
	\label{fig:V2V}
\end{figure}
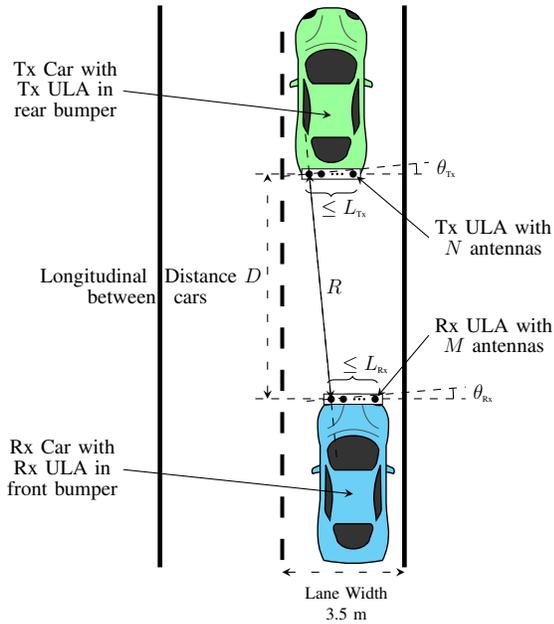

Due to the importance of V2V communication in future wireless networks, e.g. 5G, we consider the LOS MIMO design for a V2V link between two vehicles located in the same lane, where the front car (Tx car) is communicating with a rear car (Rx car) separated by a longitudinal distance $D$ as shown in Fig.~\ref{fig:V2V}. The Tx car is equipped in the \emph{rear} bumper with a Tx ULA consisting of $N$ antennas, whereas the Rx car is equipped in the \emph{front} bumper with a Rx ULA consisting of $M$ antennas. The \emph{maximum} length of the Tx and the Rx ULA is assumed to be $L_{\text{\tiny Tx}}$ and $L_{\text{\tiny Rx}}$, respectively. From Fig.~\ref{fig:V2V}, we can see that the Tx ULA and the Rx ULA are always parallel and hence, the orientation of both arrays are the same, i.e. $\theta_{\text{\tiny Tx}}=\theta_{\text{\tiny Rx}}$ (c.f. Fig.~\ref{fig:LOSMIMO}). We assume a pure LOS channel between the Tx ULA and the Rx ULA, as well as the same speed for the Tx and Rx car. We assume a carrier frequency of $f_{\text{c}}=28$ GHz ($\lambda \approx 10.7$ mm) and a normalized LOS channel as discussed in Section~\ref{Sec:ChMod}, with a fixed receive $\text{SNR}=$ 13 dB for the considered distances $D$, i.e. with perfect sidelink power control.

We consider the LOS MIMO design over a range of distances $D$ between the two cars with $10 \le D \le 100$. Due to lack of space, we do not consider the  horizontal displacement of the two cars within the lane (of width equal to $3.5$ m), which leads to slightly different orientation angles of the arrays. We assume  the cars are facing each other, such that $\theta_{\text{\tiny Tx}}=\theta_{\text{\tiny Rx}}=0$ and $R=D$. Furthermore, we assume the same number of antennas in the Tx and Rx array and set it to $3$, i.e. $N=M=3$, as well as the same antenna separation $d$ at both the Tx and Rx array, i.e. $d=d_{\text{\tiny Tx}}^{}=d_{\text{\tiny Rx}}^{}$. The maximum length of the arrays is assumed to be equal and set to $1.8$ m, in order to fit in the bumpers of a standard car, i.e. $L_{\text{\tiny Tx}}= L_{\text{\tiny Rx}} = 1.8$ m. Note that for the considered distances, $D \gg L_{\text{\tiny Tx}}= L_{\text{\tiny Rx}} = 1.8$.

From (\ref{ortho_sol_asp}) with  $d_{\text{\tiny Tx}}^{}=d_{\text{\tiny Rx}}^{}=d$, $R=D$, and $N=M=3$, the optimum antenna separation for both arrays is given by
\begin{align}
&d_{\text{}}^{} = \sqrt{ p \cdot \frac{\lambda D}{M}}  \quad \quad \text{for} \quad p \in \left\{1,2,4,5,7,8,\cdots \right\},   \label{ortho_sol_ex} 
\end{align}
where only the multiples of $M=3$ are excluded from the set of positive integers for the possible values of $p$ in \eqref{ortho_sol_asp}. To observe the multiple solutions for the optimum antenna separation $d$ which maximize the capacity, i.e. which result in an orthogonal LOS MIMO channel with $3$ equally strong eigenmodes, we plot $d$ given in (\ref{ortho_sol_ex}) as a function of the longitudinal distance $D$ between the cars for the first eight values of $p$. As mentioned before, only the solution corresponding to $p=1$ is usually considered in the literature, as this corresponds to the smallest optimum antenna separation which then results in the shortest Tx and Rx arrays.



\begin{figure}[!th]
	\begin{center}
		\begin{tikzpicture}
		\begin{axis}[ylabel= Opt. Ant. Separation $d$ (meters),
		xlabel=Distance $D$ between the cars (meters),
		grid,
		xmin=10,
		xmax=100,
		ymin=0,
		ymax=0.9,			
		xtick={10,20,...,100},
		legend pos= south east,
		]
		,  
		\foreach \F/\p/\c/\M in {1/1/myblue/o,2/2/myorange/square,3/4/myyellow/+,4/5/mypurple/diamond,5/7/mygreen/triangle,6/8/myred/asterisk,7/10/mylightblue/pentagon,8/11/magenta/x}
		{ 
			\edef\temp{\noexpand\addplot[color=\c,line width=1pt, mark=\M,] table[x index=0, y index=\F] {Ant_Spacing_vs_Distance_R_M3.txt};}
			\temp 	
			\edef\legendentry{\noexpand\addlegendentry{$p =\noexpand\pgfmathprintnumber[fixed]{\p}$}};
			\legendentry
		}
		
	
		\draw[thick, black, dashed] (axis cs: 0,0.5976) -- (axis cs: 100,0.5976);
		
		
		\node at (axis cs:70,0.6) [anchor=south west]   {\small $d=0.5976$};

		\end{axis}
		\end{tikzpicture}
	\end{center}
	\vspace{-3ex}
	\caption{Optimum Antenna Separations for the V2V link}  
	\label{fig:res1}
\end{figure}
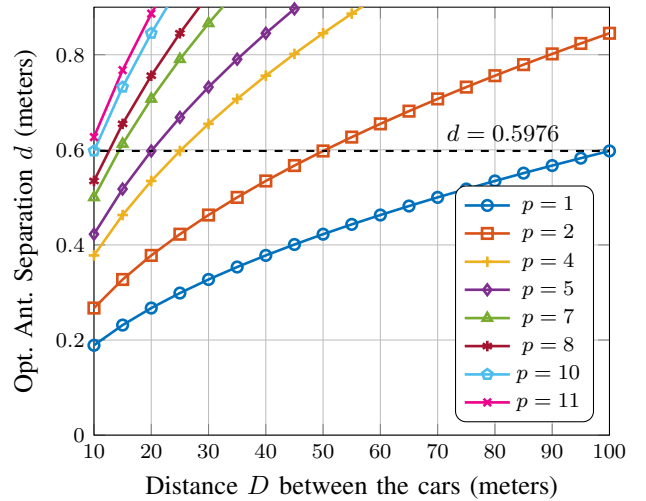

However, the curves for $p>1$ result in larger antenna separations which also maximize the channel capacity. For a given distance $D$ between the arrays, the optimum antenna separation increases with $\sqrt{p}$ as can be shown in (\ref{ortho_sol_ex}). This results in an increasing length of the arrays with $p$, given by $(M-1)  \sqrt{ p \cdot \frac{\lambda D}{M}}$. Due to the maximum length of the Tx and Rx arrays (car bumpers) in our V2V scenario given by $L_{\text{\tiny Tx}}= L_{\text{\tiny Rx}} = 1.8$ m, we consider only those solutions for $d$ which are less than or equal to $\frac{1.8}{M-1}$, i.e. with $M=3$, we consider only the optimum antenna separations which fulfill
\begin{align}
d \le 0.9 .
\end{align}
With this constraint, we observe from Fig.~\ref{fig:res1}  there are at least two possible antenna separations which guarantee a $3 \times 3$ orthogonal LOS MIMO channel for each distance $D$ in the considered range of distances up to $100$ m.

\begin{figure}[!ht]
	\begin{center}
		\begin{tikzpicture}
		\begin{axis}[ylabel= Eigenvalues of $\, \m{H}(D)\m{H}^{\text{H}}(D)$,
		xlabel=Distance $D$ between the cars (meters),
		grid,
		xmin=10,
		xmax=100,
		ymin=0,
		ymax=10,			
		ytick={0,3,6,9},
		xtick={10,20,...,100},
		legend pos= north east,
		]
		,  
		

\addplot[color=mygreen,line width=1pt,dashdotted] table[x index=0, y index=1] {Eigs_vs_Distance_R_M3sol_2_D50.txt};

\addplot[color=mylightblue,line width=1pt] table[x index=0, y index=2] {Eigs_vs_Distance_R_M3sol_2_D50.txt};

\addplot[color=myred,line width=1pt,dashed] table[x index=0, y index=3] {Eigs_vs_Distance_R_M3sol_2_D50.txt};

		
		
		
		\node at (axis cs:50,7) [anchor=south west,align=left]   {\footnotesize{$3$ eigenvalues for} \\ \footnotesize{the case $d=0.5976$}};

		\end{axis}
		\end{tikzpicture}
	\end{center}
	\vspace{-3ex}
	\caption{Eigenvalues of  $\, \m{H}(D)\m{H}^{\text{H}}(D)$ with $d=0.5976$}  
	\label{fig:res2}
\end{figure}
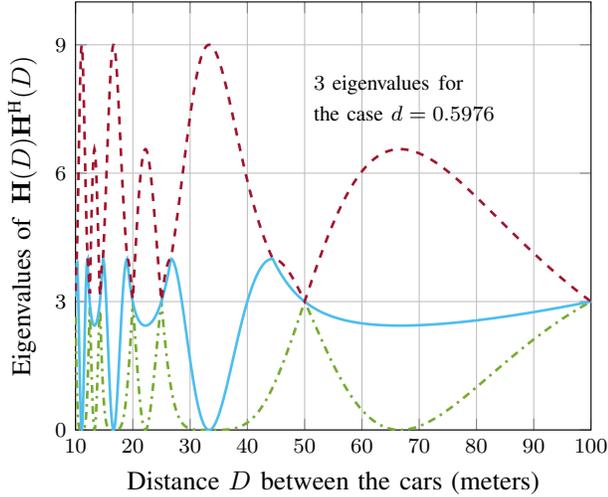

More interestingly we observe in Fig.~\ref{fig:res1} that  some antenna separations are optimum at \emph{several} distances! For example, $d=0.5976$ is an optimum antenna separation at $D~=~10,\, 12.5,\, 14.2857,\, 20,\, 25,\, 50$ and $100$ m, which can be obtained from (\ref{ortho_sol_ex}).  At these distances, the LOS MIMO channel matrix with $d=0.5976$ is orthogonal with three equally strong eigenmodes as shown in Fig.~\ref{fig:res2}, where the eigenvalues of $\m{H}(D)\m{H}^{\text{H}}(D)$ are depicted for the considered range of distances between the cars. As $\text{tr}\left(\m{H}(D)\m{H}^{\text{H}}(D)\right)=MN=9$, the capacity with $\m{H}(D)$ is maximized when the three eigenvalues of $\m{H}(D)\m{H}^{\text{H}}(D)$ are equal  to $3$. The channel matrix $\m{H}(D) \in \mathbb{C}^{3 \times 3}$ corresponds to a LOS MIMO system given by (\ref{def_H}), (\ref{def_hn}) and (\ref{def_rmn}) with a Tx and Rx ULA consisting of $3$ antennas with  an antenna separation of $0.5976$ and a distance $D$ between the Tx and Rx arrays. $\m{H}(D)$ is given as a function of $D$ to highlight its dependency on the distance $R=D$ between the arrays via (\ref{def_rmn}).  From Fig.~\ref{fig:res2}, we also see that at certain distances some eigenvalues go to zero and hence, the LOS channel $\m{H}(D)$ becomes rank deficient, e.g. at $D=34$ and $D=68$ the channel rank is $1$ and $2$, respectively.

\begin{figure}[!ht]
	\begin{center}
		\begin{tikzpicture}
		\begin{axis}[ylabel= Capacity (bps/Hz),
		xlabel=Distance $D$ between the cars (meters),
		grid,
		xmin=10,
		xmax=100,
		ymin=5,
		ymax=15,			
		ytick={1,2,...,15},
		xtick={10,20,...,100},
		legend pos= south east,
		]
		,  
		

		\addplot[color=mygreen, densely dotted, thick, line width=1pt] table[x index=0, y index=2] {Cap_vs_Distance_R_M3d05.txt};
		\addlegendentry{Max. Capacity} 
		
		\addplot[color=myblue,line width=1pt,smooth,] table[x index=0, y index=1] {Cap_vs_Distance_R_M3d05.txt};
		\addlegendentry{$d=0.5$} 
		
		
		\addplot[color=myred,line width=1pt,smooth, dashed] table[x index=0, y index=1] {Cap_vs_Distance_R_M3sol_2_D50.txt};
		\addlegendentry{$d=0.5976$} 
		
		\addplot[color=myyellow,line width=1pt,smooth,] table[x index=0, y index=1] {Cap_vs_Distance_R_M3d07.txt};
		\addlegendentry{$d=0.7$} 
		

		
		\draw[->] (axis cs: 35,6) node[right,,align=center] {{\footnotesize $2$ eigenvalues} \\ {\footnotesize go to zero}} -- (axis cs: 34,7.39) ;

		\draw[->] (axis cs: 70,9.1) node[right,align=center] {{\footnotesize $1$ eigenvalue} \\ {\footnotesize goes to zero}} -- (axis cs: 68,10.6) ;
		
		

		\node[draw,shape=circle,fill=myred,inner sep=1.5pt] at (axis cs:50,13.18) () {};
		\node[draw,shape=circle,fill=myyellow,inner sep=1.5pt] at (axis cs:68.8,13.18) () {};
		
	 	\draw[->,thick, black] (axis cs: 50,13.8) -- (axis cs: 68.8,13.8) node[above,midway,black] { \footnotesize{Shift by $1.37$}};
				
		\end{axis}
		\end{tikzpicture}
	\end{center}
	\vspace{-3ex}
	\caption{Capacity of the LOS MIMO channel for different antenna separations}  
	\label{fig:res3}
\end{figure}
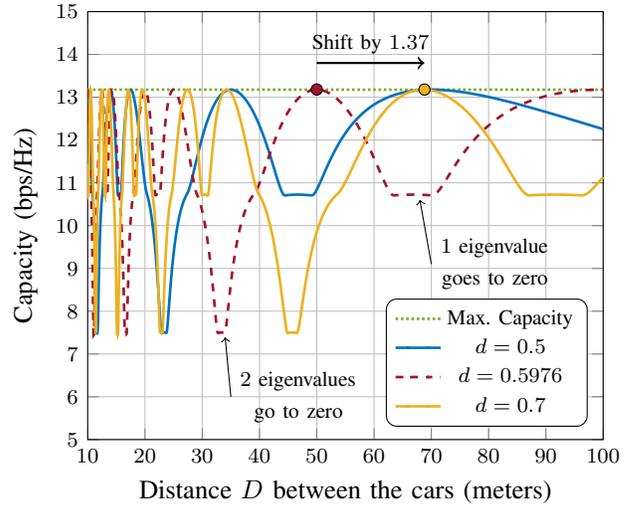

To elaborate further on the performance  over the considered range of distances, we depict in Fig.~\ref{fig:res3} the  capacity of the LOS MIMO channel for the described V2V link for three different antenna separations  
$d=0.5,\,0.5976,\,0.7$ for the Tx and Rx ULAs. In this case, the maximum capacity with an SNR of $13$ dB is $13.18$ bps/Hz, whereas the capacity  is $10.72$ and  $7.50$ bps/Hz when one or two eigenmodes go to zero, respectively. As can be seen in Fig.~\ref{fig:res3}, the maximum capacity with $d=0.5976$  is achieved for the set of distances mentioned previously. For $d=0.5$ and $d=0.7$, the maximum capacity is achieved at other sets of distances. In fact, we observe a \emph{stretching} and \emph{shift} to the right of the capacity curve as the antenna separation $d$ increases, which can be explained as follows. Given that (\ref{ortho_sol_ex}) can be rewritten as $\frac{d^{2}}{D}=p\frac{\lambda}{M}$, we can find other pairs of antenna separation $d^{\prime}$ and distance $D^{\prime}$ which achieve the same value $p \cdot\frac{\lambda}{M}$, i.e.
\begin{align}
\frac{d^{\prime,2}}{D^{\prime}} = \frac{d^{2}}{D},  \quad \quad \text{such that} \quad \quad 
D^{\prime} =  D \, \frac{d^{\prime,2}}{d^{2}}.   
\label{Ddexpression}
\end{align}
For instance, from Fig.~\ref{fig:res1} the optimum antenna separation for $p=2$ at $D=50$  is $d=0.5976$. From (\ref{Ddexpression}), the distance $D^{\prime}$ which achieves the same value $p \cdot\frac{\lambda}{M}$ as the previous setting but with $d^{\prime}=0.7$ is given by $D^{\prime}=50\cdot\frac{0.7^2}{0.5976^2}=68.8$ m. Thus, in Fig.~\ref{fig:res3} the point on the capacity curve for $d=0.5976$ at $D=50$ is shifted to the right by a factor of $\frac{0.7^2}{0.5976^2} \approx 1.37$ when the antenna separation $d=0.7$ is employed. The stretching of the capacity curve can also be explained in a similar manner.

\section{Conclusion}
\label{Sec:Con}
We have derived the general expression for the optimum antenna separation product for maximizing the capacity of a LOS MIMO channel with a Tx and Rx ULA. The expression leads to multiple solutions of the optimum antenna separation product which depend on the number of Tx and Rx antennas. We have proposed to exploit the multiple solutions for the LOS MIMO design over a range of distances between the transmitter and receiver, such as for V2V. We have shown that larger antenna separations can be beneficial and that some antenna separations are optimum at several distances. The provided results can serve as guidelines for the LOS MIMO design for V2V. Future work includes considering non-uniform linear arrays as well as the ground reflection in the V2V link. 

\section*{Acknowledgment}
{The authors would like to acknowledge support of this work
under the 5GPPP European Project 5GCAR (grant agreement
number 761510).}



%

\end{document}